\documentclass[10pt,twocolumn,letterpaper]{article}

\usepackage{cvpr}              

\usepackage{times}
\usepackage{helvet}
\usepackage{courier}
\usepackage{natbib}
\usepackage[accsupp]{axessibility} 
\usepackage{amsmath}
\usepackage{graphicx}
\usepackage{multirow}
\usepackage{multicol}
\usepackage{booktabs}
\usepackage{array}
\usepackage{colortbl}
\usepackage{makecell}
\usepackage{xcolor}
\usepackage{caption}
\usepackage{enumitem}
\usepackage{csquotes}
\usepackage{url}
\usepackage{algorithm}
\usepackage{algorithmic}
\usepackage{newfloat}
\usepackage{listings}
\usepackage{pifont}

\definecolor{cvprblue}{rgb}{0.21,0.49,0.74}
\usepackage[pagebackref,breaklinks,colorlinks,allcolors=cvprblue]{hyperref}

\def\paperID{9145}
\def\confName{CVPR}
\def\confYear{2026}

\title{ADSeeker: A Knowledge-Grounded Reasoning Framework for \\ Industry Anomaly Detection and Reasoning}

\author{
Kai Zhang\textsuperscript{1} \quad
Zekai Zhang\textsuperscript{1} \quad
Xihe Sun\textsuperscript{1} \quad
Anpeng Wang\textsuperscript{1} \quad
Jingmeng Nie\textsuperscript{1}\\[0.3em]
Qinghui Chen\textsuperscript{1} \quad
Han Hao\textsuperscript{1} \quad
Jianyuan Guo\textsuperscript{2} \quad
Jinglin Zhang\textsuperscript{1}$^\dagger$\\[0.5em]
\textsuperscript{1}Shandong University \quad
\textsuperscript{2}City University of Hong Kong\\[0.3em]
{\tt\small 202200171008@mail.sdu.edu.cn, jinglin.zhang@sdu.edu.cn}\\[0.2em]
{\small $^\dagger$Corresponding author}
}

\begin{document}
\maketitle
\begin{abstract}

Automatic vision inspection holds significant importance in industry inspection. While multimodal large language models (MLLMs) exhibit strong language understanding capabilities and hold promise for this task, their performance remains significantly inferior to that of human experts. In this context, we identify two key challenges: (i) insufficient integration of anomaly detection (AD) knowledge during pre-training, and (ii) the lack of technically precise and context-aware language generation for anomaly reasoning. To address these issues, we propose \textbf{ADSeeker}, an anomaly task assistant designed to enhance inspection performance through knowledge-grounded reasoning. ADSeeker first leverages a curated visual document knowledge base, SEEK-MVTec\&VisA (\textbf{SEEK-M\&V}), which we construct to address the limitations of existing resources that rely solely on unstructured text. SEEK-M\&V includes semantic-rich descriptions and image-document pairs, enabling more comprehensive anomaly understanding. To effectively retrieve and utilize this knowledge, we introduce the Query Image-Knowledge Retrieval-Augmented Generation (\textbf{Q2K RAG}) framework. To further enhance the performance in zero-shot anomaly detection (ZSAD), ADSeeker leverages the \emph{Hierarchical Sparse Prompt} mechanism and type-level features to efficiently extract anomaly patterns. Furthermore, to tackle the challenge of limited industry anomaly detection (IAD) data, we introduce the largest-scale AD dataset, Multi-type Anomaly (\textbf{MulA}), encompassing 72 multi-scale defect types across 26 categories. Extensive experiments show that our plug-and-play framework, ADSeeker, achieves state-of-the-art zero-shot performance on several benchmark datasets.

\end{abstract}    
\section{Introduction}

\begin{figure}
    
    \centering
    \includegraphics[width=1\linewidth]{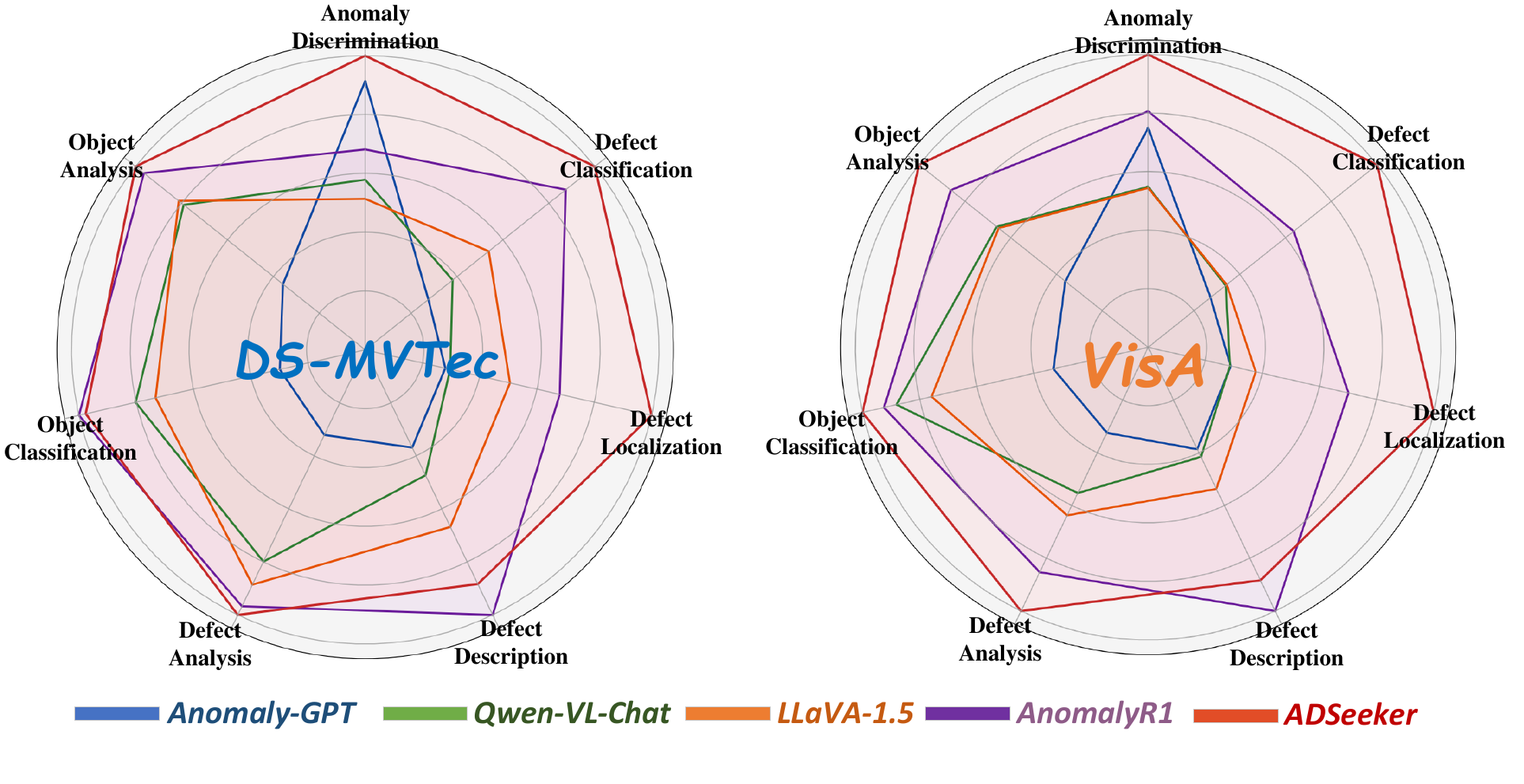}
    \vspace{-14pt}
    \caption{\textbf{Visualization of the reasoning capability.} Comparative results of ADSeeker against state-of-the-art models on the MMAD benchmark. Notably, the framework is specifically designed for IAD, its performance on defect analysis and localization tasks is better than any other IAD models.}
    \label{fig1}
\end{figure}

\begin{figure*}
    \centering
    \includegraphics[width=1\linewidth]{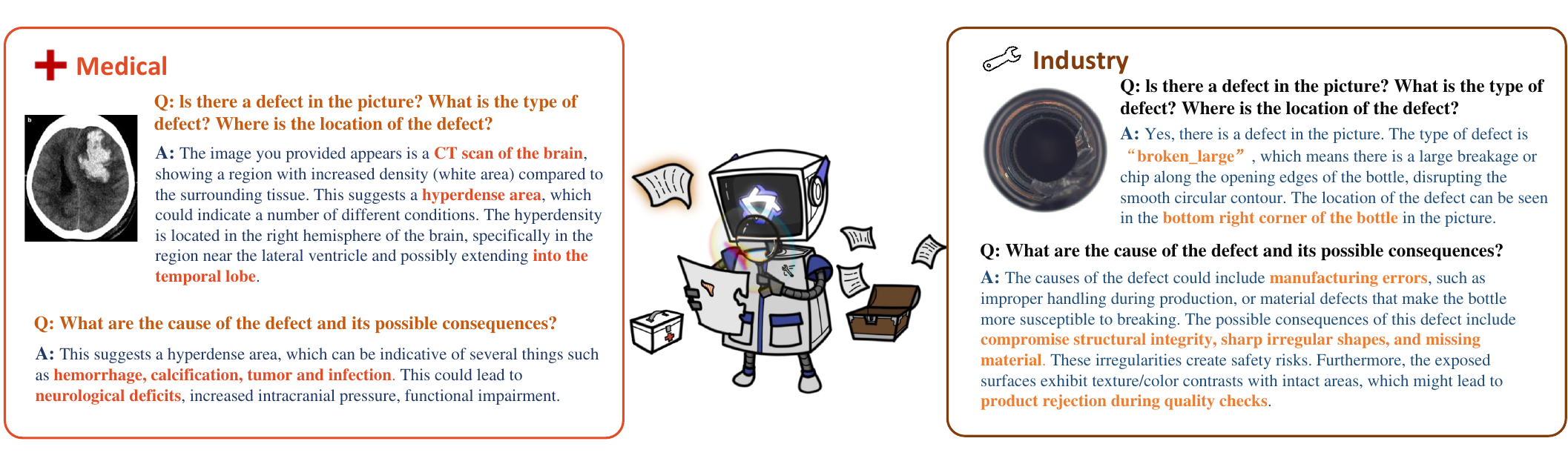}
    \vspace{-14pt}
    \caption{\textbf{Industrial and medical image anomaly reasoning results of ADSeeker.} The results include accurate anomaly location, anomaly description, defect classification, and other fine-grained information.}
    \label{fig2}
\end{figure*}

Multimodal Large Language Models (MLLMs), such as GPT-4 and LLaVA \cite{Li2024LLaVAOneVisionEV}, have shown significant promise across various intellectual human tasks, including writing, programming, and Industry Anomaly Detection (IAD). In IAD tasks, anomaly reasoning is distinct from anomaly detection. Anomaly reasoning necessitates a finer level of granularity and logical reasoning in the model's responses. This distinction is vital in identifying the roles of a quality inspector versus a domain expert. Recent studies have explored the use of MLLMs for IAD tasks, including detection, description, analysis, etc., by training models such as AnomalyGPT \cite{gu2023anomalygptdetectingindustrialanomalies} and Anomaly-OV \cite{xu2025zeroshotanomalydetectionreasoning}, to incorporate domain-specific knowledge for improved anomaly reasoning. While some models show promise in Zero-Shot Anomaly Detection (ZSAD), they often falter in accurate defect analysis and detailed descriptions. Moreover, standard fine-tuning risks catastrophic forgetting of foundational alignments, compounded by limited datasets that cause overfitting to narrow answer templates and poor generalization, rather than robust context-aware anomaly reasoning. Tracing these performance bottlenecks to their core, we identify two key challenges that existing methods\cite{Zhang2024RepresentationLB} have yet to fully address: i) insufficient integration of anomaly detection knowledge during pre-training, hindering precise localization, and ii) a lack of technically precise and context-aware language generation for anomaly reasoning. To tackle these issues, we propose \textbf{ADSeeker}, which addresses the limitations of domain knowledge through the Query Image—Knowledge Retrieval-Augmented-Generation (Q2K RAG).

The Q2K RAG module achieves multimodal hybrid retrieval, seeking the knowledge document most relevant to the query image. Specifically, the query image and knowledge document are encoded into the \emph{Key Embedding} and the \emph{Lock Embedding} respectively. Subsequent multimodal retrieval is performed in the joint feature space between these representations. To compensate for the lack of knowledge base in IAD domain, we establish the first-ever knowledge base SEEK-M\&V composed of visually rich industry anomaly documents. In SEEK-M\&V, each knowledge document has specific reference pages containing the defect type and the defect analysis. Moreover, we utilize DeepSeek-R1 \cite{deepseekai2025deepseekr1incentivizingreasoningcapability} to generate semantic-rich descriptions and application scenario introductions to expand the content of SEEK-M\&V. As the saying goes, \enquote{a picture is worth a thousand words.} SEEK-M\&V enriches knowledge representation by preserving graphic information for each anomaly type, offering a valuable reference for anomaly reasoning. Compared to existing training-based approaches\cite{Chen2025DistilledLL}, it significantly improves both efficiency and effectiveness.

Existing approaches\cite{Li2025HGCFHG} have introduced learnable prompts to adapt models for ZSAD. However, object-level prompts (\emph{e.g.}, identifying objects such as ``wood" or ``leather") are incapable of providing detailed anomaly patterns, and publicly available AD datasets struggle with semantically aligning these patterns. To address these gaps, we introduce Multi-type Anomaly (MulA), the largest-scale dataset for ZSAD, which includes 72 defect types across 8 scenarios. Additionally, ADSeeker utilizes a \emph{Hierarchical Sparse
Prompt} (HSP) mechanism to extract type-level features (\emph{e.g.}, distinguishing between ``scratch" or ``hole" anomalies) and reduce hallucinations caused by suspected anomaly regions.

Extensive experiments validate the effectiveness and efficiency of ADSeeker. For example, anomaly reasoning experiments on MMAD \cite{Jiang2024MMADAC} benchmark demonstrate that ADSeeker outperforms other general models. Additionally, experiments on 12 datasets from both industrial and medical sectors highlight ADSeeker's superiority in ZSAD tasks. To validate the effectiveness of type-level features, we performed extensive tests to visualize the impact of type-level features across a range of products. We also conducted ablation studies and evaluated the performance-resource trade-offs within our ADSeeker.

Our main contributions are summarized as follows:
\begin{itemize}
    \item We propose ADSeeker, a novel framework that utilizes Q2K RAG and AD Expert module for IAD tasks, enhancing specialization and fine granularity in reasoning.  
    \item A \emph{Hierarchical Sparse Prompt} (HSP) mechanism is introduced to extract general anomaly patterns, resulting in substantial improvements in ZSAD performance.
    \item We propose SEEK-M\&V, the first visual document knowledge base for anomaly reasoning. In addition, we introduce MulA, the largest-scale AD dataset to date, which addresses the challenge of limited IAD data the lack of type-level annotations.
\end{itemize}
\begin{figure*}
    \includegraphics[width=1.00\linewidth]{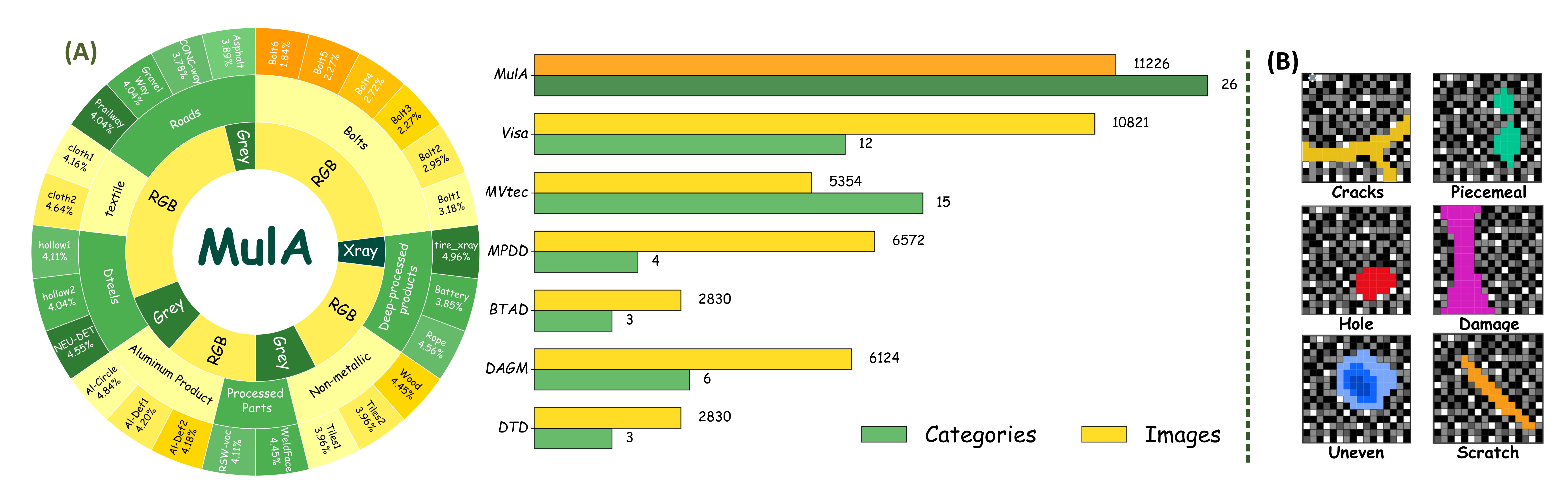}
    \captionof{figure}{\textbf{Details of MulA Dataset.} (A) Statistics and components of the MulA Dataset, the  newest dataset with the largest number and most types, and (B) examples of different types of type-level feature. Our dataset can also be organized by its defect type to enhance the Anomaly Detection performance of our ADSeeker through clustering.}
    \label{fig3}
\end{figure*}
\section{Related Work}
\textbf{MLLM for industry anomaly detection (IAD).}
 Recent advances have demonstrated the potential of MLLMs in IAD. For example, AnomalyGPT is directly fine-tuned  on IAD datasets to inject domain knowledge into the base model. Moreover, to effectively extract detailed feature from abnormal samples, Myriad \cite{Li2023MyriadLM} developed an Expert Perception module, which is designed to harness the prior knowledge of vision experts. However, these models are constrained by the scope of their training knowledge and exhibit limited domain generalization. Existing IAD approaches\cite{Zhang2025ZeroShotLI} rarely become a trustworthy automatic vision inspection assistant \cite{,Yao2024MiniCPMVAG,Chen2023InternVS,Liu2023ImprovedBW,liu2024llavanext}. In contrast, building on the excellence of our ADSeeker framework, the anomaly reasoning performance of base models have been notably enhanced, as shown in \cref{tab4}.

\noindent\textbf{Zero-Shot Anomaly Detection (AD).} 
Recently, WinCLIP \cite{jeong2023winclipzerofewshotanomalyclassification} makes an early attempt at introducing VLMs into ZSAD tasks, which leverage CLIP \cite{radford2021learningtransferablevisualmodels} to compute the similarities between patch embeddings and textual embeddings. On this basis, AnomalyCLIP \cite{zhou2024anomalyclipobjectagnosticpromptlearning} enhances the ZSAD performance of CLIP by proposing object-agnostic prompt learning. To tackle the issue of non-adaptation to common anomaly patterns, AdaCLIP \cite{Cao_2024} proposes hybrid dynamic learnable prompts which could flexibly adapt to the query images. However, existing ZSAD methods \cite{gu2024filozeroshotanomalydetection,Cao2023,Oquab2023DINOv2LR,Kirillov2023SegmentA} cannot extract the generalizable anomaly patterns due to the object-level prompts. In contrast, our proposed HSP module could extract detailed features from the query image and learn semantic-rich information from type-level features, enhancing its sensitivity to anomalous features at \cref{fig3}(B). 
\section{Datasets and Knowledge Base}

\subsection{MulA: Multi-type Anomaly Dataset}
 Existing AD datasets mainly focus on providing object-level (\emph{e.g.}, ``wood", ``leather", ``plastic") information \cite{MVTec,VisA,MPDD,Head-CT-hemorrhage}, while the type-level feature (\emph{e.g.}, ``crack", ``hole", ``scratch") plays an essential role in extracting defect-region features from the query image. To this end, we propose Multi-type Anomaly dataset (MulA), which contains 11,226 images across 26 categories, surpassing existing AD datasets in scale and diversity. We collect the data from different scenarios, including industry, ordinary, and workshops, bringing multi-type generalizable anomaly patterns. 

We constructed a high-quality training dataset through manual mask annotation. This dataset addresses two primary challenges: insufficient defect type diversity and a scarcity of normal (defect-free) samples. To enrich the defect distribution, we introduced Gaussian noise. Furthermore, to mitigate overfitting and poor ZSAD generalization caused by replicating limited normal samples, we generated a diverse set of normal samples by applying image composition and geometric transformations to existing anomaly data. As shown in \cref{fig3}(A) , MulA incorporates data across various scales, encompassing grayscale, RGB, and X-ray, exhibiting a wider generalization in industry. Besides the treatment of samples, we categorize industrial scenarios according to industry types. Furthermore, we compared classical datasets from both medical and industrial fields, as depicted in \cref{fig3}(A).

\subsection{SEEK-M\&V: IAD Knowledge Base}

Over the years, there is a lack of knowledge base for RAG frameworks in the field of anomaly detection \cite{Zhao2024RetrievalAugmentedGF, Wang2025ViDoRAGVD, Wu2025VisualRAGBT, Lee2025RAGEnhancedCL}. Moreover, current knowledge base tends to focus on textual information, while image features play an essential role in anomaly detection \cite{Yu2024VisRAGVR}. Building upon these insights, we build the first visual document knowledge base SEEK-M\&V in IAD tasks.

\begin{figure*}
    \centering
    \includegraphics[width=1.0\linewidth]{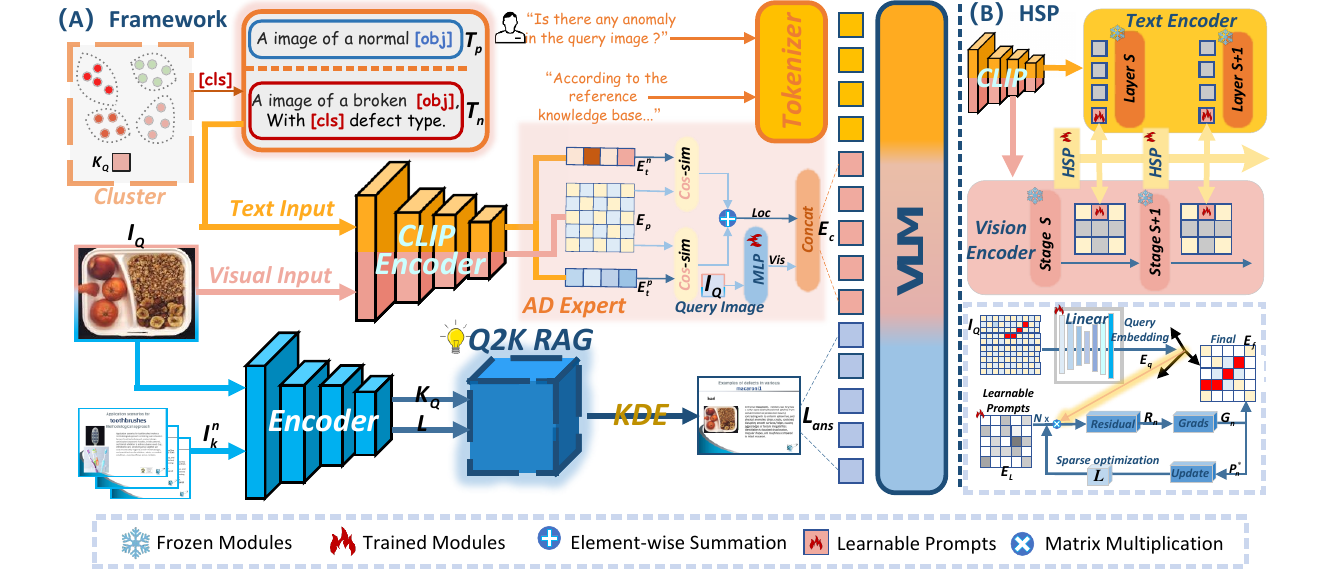}
    \caption{\textbf{The architecture of ADSeeker.} It consists of two main knowledge-infused pathways: (1) The query image and knowledge documents are fed to the Q2K RAG to retrieve high-relevant domain knowledge. (2) The AD Expert integrates defect-region information and type-level features into semantic-rich visual tokens which will be passed into VLM.}
    \label{fig4}
\end{figure*}

To enhance the specialization of domain knowledge in the SEEK-M\&V, we present detailed information about the anomaly with the official documentation for AD datasets, such as anomaly type, anomaly analysis, and anomaly description. Meanwhile, we find production scenarios and application scenarios for various products to enrich the background information. Our SEEK-M\&V contains both textual expertise and visual features. With generalizable forms, multimodal features and semantic-rich information, our SEEK-M\&V  has great potential in generalizing to other domains. 
These inherent strengths make it an effective solution to cross-disciplinary problems.

\section{Method}
In this section, we describe our proposed ADSeeker framework for anomaly detection and reasoning tasks, as shown in \cref{fig4}. Recent studies reveal that pretrained MLLMs function as generalists, possessing a broad knowledge but underperforming in specialized domains, such as IAD. Therefore, our goal is to introduce a knowledge-injected framework designed to equip the generalist with the AD domain knowledge. This approach circumvents the need for pre-training while preserving the generalization capacity of the original model. We utilize the publicly available MLLM (Qwen2.5-VL \cite{Bai2025Qwen25VLTR} ) as the backbone of ADSeeker and the CLIP model (VIT-L/14@336px) as the encoder of the AD Expert module. The CLIP encoder aligns the visual information and the textual information to perform anomaly detection, and the VLM is responsible for textual instruction processing and complex reasoning.

\subsection{Framework Overview}
The framework of ADSeeker is illustrated in \cref{fig4}(A).  We store the type-level feature in learnable textual prompts $\left(T_p,T_n\right )$, such as \enquote{A image with [\textbf{cls}] defect type}, which will be fed to CLIP Text Encoder. We leverage the CLIP encoder to align the semantic-visual features. In the process of feature fusion, the visual embedding $E_{p}$ and textual embedding $E_{t}$ extracted by the CLIP Encoder will be fed to AD Expert module and then integrated into the anomaly prior feature $E_{c}$. Moreover, we introduce the HSP mechanism to preserve crucial visual features in CLIP. Specifically, learnable embeddings are leveraged to sparsify invalid information from query images layer by layer, while they are combined with standard tokens at the beginning of each stage.

In the process of knowledge retrieval, the query image $I_{q}$ and knowledge base $I_{k}^n$ will be processed by the encoder to generate key feature $K_Q$ and lock feature $L$:
\begin{equation}
{L}=\left \lbrace L_0,L_1,L_2,...,L_{n-1}\right \rbrace.
\end{equation}
The Q2K RAG module utilizes the key feature $K_Q$ to retrieve the most relevant domain knowledge \textbf{$L_{ans}$}. With the injection of \textbf{$L_{ans}$} and anomaly prior feature $E_{c}$, ADSeeker could achieve remarkable performance on anomaly reasoning.  

The key features of MulA datasets undergo centroid-based clustering aligned with predefined cluster centers. Subsequently, key feature $K_Q$ is matched to the optimal cluster centroid to generate learnable textual prompts $T_p,T_n$. With the type-level features of MulA dataset, our model could automatically generate accurate prompts and understand common defects among abnormal samples. Compared to commonly used textual prompts in ZSAD tasks, like \enquote{ Img of a [\textit{\textbf{obj}}]}, type-level prompts are generally more detailed and versatile in scope, as they encompass condensed content.

\subsection{AD Expert:Hybrid Visual Fusion}
When performing \textbf{IAD} tasks, the original MLLMs process basic Visual tokens without additional prior knowledge, such as localization and discriminatory information. To utilize anomaly prior features in the inference stage, we introduce the AD Expert module that transforms fine-grained information from query image and textual prompts into visual tokens. 

We derive the localization and discriminatory information by calculating the cosine similarities between patch embeddings $E_p$, and positive textual embeddings $E_t^p$ and negative textual embeddings $E_t^n$. We define the anomaly localization information $Loc\in{R}^{x_1\times d}$ as follows:
\begin{equation}
Loc=\textit{Unsample}(\frac{\textit{cos}(E_p,E_t^n)}{\textit{cos}(E_p,E_t^p)+\textit{cos}(E_p,E_t^n)}).
\end{equation}

As illustrated in \cref{fig4}, we set a neural network to convert query image $I_Q$ into visual embeddings $Vis\in{R}^{x_2\times d}$, which will be integrated with the anomaly information to form anomaly prior embeddings $E_{c}={\{Loc,Vis\}}$. Given the semantic-rich visual embeddings, the defect characteristics will help the model generate fine-grained anomaly descriptions and enhance the image-level anomaly detection performance.    

\begin{figure}[t]
\centering
    \includegraphics[width=\linewidth]{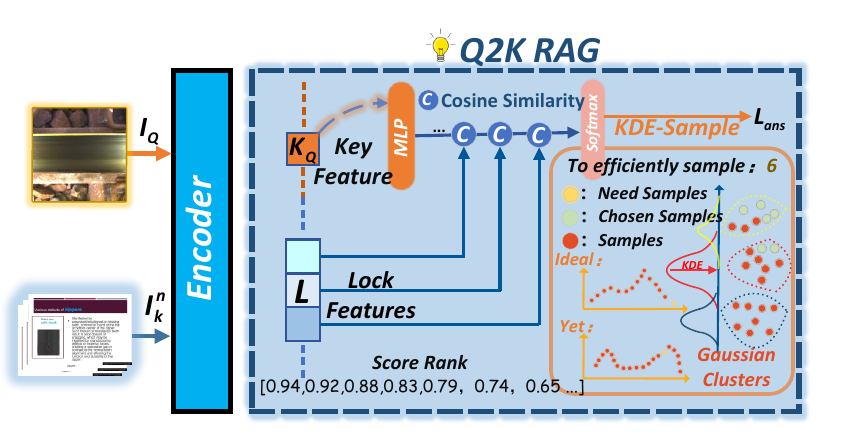}
\caption{\textbf{Illustration of the core architecture of Q2K RAG.} Solving the problem of repetitive retrieval in industrial knowledge base through KDE-Sample strategy.}
\label{fig5}
\end{figure}

\subsection{Q2K RAG:KDE-Sampled Multimodal RAG}

Instead of utilizing textual domain knowledge, our framework leverages multimodal documents in order to equip ADSeeker with AD domain knowledge. The lightweight pre-trained Q2K RAG aligns the projection spaces $\mathcal{M}=\{ K_Q,L \}$ of key feature and lock features. We adopt a similar methodology to VisRAG. To measure the relevance between the query image and the multimodal documents, we compute the similarity $\mathcal{S}$ of the embeddings: 
\begin{equation}
\mathcal{S}\left ({K_Q},L\right )=\left \lbrace S_i|cos(K_Q,L_i), L_i\in L \right \rbrace.
\end{equation}

Then the multimodal similarity $\mathcal{S}$ will be sorted to seek for the required domain knowledge $L_ans$. As shown in \cref{fig5}, in an ideal situation, the most relevant knowledge is most inspiring for the problem. We just need to retrieve the $\text{top-K}\downarrow$ expertise. Extensive experiments demonstrate that our seek accuracy rate has reached 83\%. However, there is still some possibility that required domain knowledge $L_ans$ is only at a relatively high rank, which greatly increases the difficulty of multimodal retrieval. Furthermore, SEEK-M\&V is a knowledge base organized by categories, where different defect types of the same kind of object have a high degree of similarity. The simple top-k strategy has difficulty in retrieving product-related scenario information.

To tackle these issues, we introduce the KDE-Sample to seek unevenly distributed domain knowledge. Interestingly, the similarities of the same information across categories (defect, workshops, etc ) basically obey Gaussian distribution:
\begin{equation}
N=\left \lbrace  I_n\sim\mathcal{G}(\mu_n,\sigma_n^2) |I_n \in \mathcal{S}\right \rbrace,
\end{equation}
where $\mathcal{G}$ represents a Gaussian distribution, with $\mu_n$,$\sigma_n^2$ indicating the $n_{th}$ mean and variance. Given the sorted similarity scores $\mathcal{S}$, we leverage the Bayesian Gaussian Mixture Modeling to infer the best class number K and perform GMM clustering for similarity scores:
\begin{equation}
GMM(\mathcal{S})=\left [ i_1, i_2...i_k |i\sim \mathcal{G}(\mu,\sigma^2) \right ].
\end{equation}
After that, we calculate the probability density weights $W^*$ through the kernel density estimate (KDE) algorithm. In order to quantify the quality score of each cluster, we deflate the original distribution:
\begin{equation}
W_n=\frac{1}{K}\sum_{m\in i_n}^{}\text{exp}(W^*(m)-\max_nW^*(n)). 
\end{equation}
During the subsequent dynamic retrieval process, we sort the samples in descending order of score for each cluster and dynamically select the samples from each cluster based on the ratio $W_n$. After the selected individual reaches the sampling threshold, we terminate the selection process. The textual and visual retrieval pipelines demonstrate varying levels of performance for different features. Without adaptive sampling, the combined retrieval can become redundant. KDE-Sample strategy optimizes performance relative to data distribution, underscoring the value of information value in hybrid retrieval. 

\subsection{Hierarchical Sparse Prompt}
\begin{table*}

\centering
\setlength{\tabcolsep}{7pt} 
\begin{tabular}{>{\raggedright\arraybackslash}m{2.1cm} *{9}{>{\centering\arraybackslash}p{1.3cm}}} 
\toprule
\multicolumn{1}{c}{} & \multicolumn{4}{c}{Industrial Defects} & \multicolumn{3}{c}{Medical Anomalies} & \multicolumn{1}{c}{} \\ 
\cmidrule(lr){2-5}\cmidrule(lr){6-8}
\multicolumn{0}{c}{\multirow{-2}{*}{\centering Model}} & MVTecAD & VisA & BTAD & MPDD & BrainMRI & HeadCT & Br35H & \multirow{-2}{*}{Average} \\
\hline
CLIP\cite{Ramesh2022HierarchicalTI}        & 74.1 & 66.4 & 34.5 & 54.3 & 73.9 & 56.5 & 78.4 & 62.6 \\
WinCLIP\cite{jeong2023winclipzerofewshotanomalyclassification}    & 91.8 & 78.8 & 68.2 & 63.6 & 92.6 & 90.0 & 80.5 & 80.8 \\
AnomalyCLIP\cite{zhou2024anomalyclipobjectagnosticpromptlearning} & 91.5 & 82.1 & 88.3 & 77.0 & 90.3 & 93.4 & 94.6 & 88.2 \\
AdaCLIP\cite{Cao_2024}    & 89.2 & 85.8 & 88.6 & 76.0 & \underline{94.8} & 91.4 & \underline{97.7} & 89.1 \\
AnomalyOV\cite{xu2025zeroshotanomalydetectionreasoning}  & \underline{94.0} & \underline{91.1} & \underline{89.0} & 81.7 & 93.9 & \textbf{97.6} & 95.5 & \underline{91.8} \\
\rowcolor{gray!25} \textbf{ADSeeker}    & \textbf{94.3} & \textbf{91.5} & \textbf{94.0} & \textbf{85.9} & \textbf{97.5} & \underline{96.6} & \textbf{97.9} & \textbf{94.0} \\
\bottomrule
\end{tabular}
\caption{\textbf{Image-Level AUROC performance of existing ZSAD approaches.} ADSeeker surpasses most AD expert models on this subtask. The top-performing result is presented in bold, while the second-best result is underlined.}
\label{Tab1}
\vspace{-3mm}
\end{table*}

Drawing inspiration from the attention distribution of human industry inspection, that defect region of query images deserves the majority of attention rather than distributed evenly, we design a learnable hierarchical sparse prompt (HSP) module that integrates compression awareness and sparse optimization. The HSP module also addresses issues of computational inefficiency and vulnerability to noise. The design of this module is primarily affected by AnomalyCLIP, AdaCLIP, and AnomalyGPT.

As illustrated in \cref{fig4}(B), given the query image $I_{Q}$, the defect features of query image are passed through linear layers to obtain the query embeddings $E_q\in{R}^{B\times d}$. Then we introduce learnable prompt embeddings $E_l\in{R}^{K\times d}$ and adaptive parameters $P_i\in{R}^{B\times K}$ which are then fed into the sparse optimization module. The prompt embeddings $E_l$ extract essential features from the embeddings $E_q$ through $\mathcal{N}$ iterative updates. At the $n_{th}$ round of iteration, the current residual $R_n$ can be obtained by: \begin{equation}
R_n=E_q-P_nE_l^{n-1},
\end{equation}
the update gradient $G_n$ can be represented by \textit{$G_n=-\mu\cdot E_q^TP_nE_l$}, where $\mu$ represents a pre-defined descent rate. To ensure the best match between the final embeddings $E_l^n$ and the fundamental feature, we calculate the iterated learnable conversion factor $P_n^{*}$ by combining Iterative Soft Thresholding Algorithm and sparsification theory:

\begin{equation}
\scalebox{0.75}{$\displaystyle 
P_n^{*} = \text{\textit{Sign}}\left(P_n - \frac{G_n}{\sigma_{\max}(E_l^T E_l)} \right)\left( \left | P_n - \frac{G_n}{\sigma_{\max}(E_l^T E_l)} \right|-\frac{\lambda}{G_n} \right)\cdot E_l
$}
\end{equation}

At the end of each iteration, the learnable embeddings will be automatically updated by back propagation. Specifically, the variation of parameters is based on the loss function $\mathcal{L}$ between the learnable embeddings and the query embeddings, which can be expressed as:
\begin{equation}
\mathcal{L}=\min_P\frac{1}{2}\|E_q-P_n^*E_l\|_2^2+\lambda\|P_n^*\|_1,
\end{equation}
the sparsity coefficient \textit{$\lambda$} in the above formula represents the drop ratio of the fundamental features of query image. This mechanism for sparse optimization is designed to accommodate features derived from linear layers, yet the actual features extracted remain totally blind to us. To tackle this issue, we pre-train the network for feature extraction in stages to enhance its sensitivity to defect regions. The linear layers are trained to enhance sensitivity to defect regions.

The final embeddings $E_f$ are integrated to a certain depth during the forward pass in the vision and text encoder. As we explained in the \cref{sec4.1}, the type-level textual prompts will be concatenated with defect-region features to enhance ADSeeker's fine-grained understanding of the anomaly pattern. Extensive experiments have demonstrated the potential of this module in bridging the gap between human anomaly inspector and ADSeeker.

\begin{table*}[t!]
\centering
\setlength\tabcolsep{3pt}
\renewcommand{\arraystretch}{1}
\scalebox{0.73}{ 
\begin{tabular}{c|c|c|c|cccc|cc|c}
\toprule
\multirow{2.5}{*}{\textbf{Model}}& 
\multirow{2.5}{*}{\textbf{Scale}}& 
\multirow{2.5}{*}{\textbf{Setting}}& 
\multicolumn{1}{c|}{\textbf{Anomaly}}& 
\multicolumn{4}{c|}{\textbf{Defect}} & 
\multicolumn{2}{c|}{\textbf{Object}} & 
\multirow{2.5}{*}{\textbf{Average}}\\
\cmidrule(lr){4-8} \cmidrule(lr){9-10} 
 & & & \textbf{Discrimination}& 
\textbf{Classification}& \textbf{Localization}& \textbf{Description}& \textbf{Analysis}& 
\textbf{Classification}& \textbf{Analysis}& \\
\midrule
\multirow{2}{*}{LLaVA-OV\cite{Li2024LLaVAOneVisionEV}}& \multirow{2}{*}{7B} & Baseline & 51.17 & 46.13 & 41.85 & 62.19 & 69.73 & 90.31 & 80.93 & 63.19 \\
\cmidrule(lr){3-11}&  & Seek-Setting & 55.35 {\color{black}(+4.18)} & 49.86 {\color{black}(+3.73)} & 54.34 {\color{black}(+12.49)} & 57.17 {\color{gray}(-5.02)} & 73.20 {\color{black}(+3.47)} & 92.07 {\color{black}(+1.76)} & 84.31 {\color{black}(+3.38)} & 66.61{\color{black}(+3.42)} \\
\midrule
\multirow{2}{*}{LLaVA-NeXT\cite{liu2024llavanext}}& \multirow{2}{*}{7B} & Baseline & 57.64 & 33.79 & 47.72 & 51.84 & 67.93 & 81.39 & 74.91 & 59.32 \\
\cmidrule(lr){3-11}
 &  & Seek-Setting & 64.10 {\color{black}(+6.46)} & 53.29 {\color{black}(+19.50)} & 58.13 {\color{black}(+10.41)} & 54.09 {\color{black}(+2.25)} & 71.20 {\color{black}(+3.27)} & 87.70 {\color{black}(+6.31)} & 73.20 {\color{gray}(-1.71)} & 65.96{\color{black}(+6.64)} \\
\midrule
\multirow{2}{*}{InternVL2\cite{Chen2023InternVS}}& \multirow{2}{*}{8B} & Baseline & 59.97 & 43.85 & 47.91 & 57.60 & 78.10 & 74.18 & 80.37 & 63.14 \\
\cmidrule(lr){3-11}
 &  & Seek-Setting & 63.26 {\color{black}(+3.29)} & 58.71 {\color{black}(+14.86)} & 49.45 {\color{black}(+1.54)} & 56.45 {\color{gray}(-1.15)} & 82.73 {\color{black}(+4.63)} & 84.79 {\color{black}(+10.61)} & 80.93 {\color{black}(+0.56)} & 68.05{\color{black}(+4.91)} \\
\midrule
\multirow{2}{*}{Qwen2.5-VL\cite{Bai2025Qwen25VLTR}}& \multirow{2}{*}{3B} & Baseline & 51.10 & 41.07 & 44.87 & 61.43 & 75.99 & 86.54 & 79.58 & 62.94 \\
\cmidrule(lr){3-11}&  & Seek-Setting & 58.17 {\color{black}(+7.07)} & 48.71 {\color{black}(+7.64)} & 53.67 {\color{black}(+8.80)} & 69.79 {\color{black}(+8.36)} & 80.36 {\color{black}(+4.37)} & 86.78 {\color{black}(+0.24)} & 82.25 {\color{black}(+2.67)} & 68.53{\color{black}(+5.59)} \\
\midrule
\multirow{2}{*}{Qwen2.5-VL\cite{Bai2025Qwen25VLTR}}& \multirow{2}{*}{7B} & Baseline & 55.35 & 49.86 & 54.34 & 57.18 & 73.20 & 92.07 & 84.31 & 66.62 \\
\cmidrule(lr){3-11}&  & Seek-Setting & 68.62 {\color{black}(+13.27)} & 53.82 {\color{black}(+3.96)} & 55.57 {\color{black}(+1.23)} & 60.51 {\color{black}(+3.33)} & 77.40 {\color{black}(+4.20)} & 89.91 {\color{gray}(-2.16)} & 83.48 {\color{gray}(-0.83)} & 69.90{\color{black}(+3.28)} \\
\bottomrule
\end{tabular}
}
\caption{\textbf{Anomaly Reasoning Performance.} Performance comparison of both ADSeeker (Qwen2.5-VL under Seek-Setting) and other open-source MLLMs in MMAD with the standard 1-shot setting. Notably, each model is also tested under Seek-Setting.}
\label{tab4}
\end{table*}

\section{Experiments}

\subsection{Zero-shot Anomaly Detection}
\label{sec4.1}
 We conduct extensive experiments on publicly available AD benchmarks from industrial and medical domains. Specifically, for the industry domain, we utilize MVTec AD\cite{MVTec}, VisA\cite{VisA}, BTAD\cite{Mishra2021VTADLAV}, and MPDD\cite{MPDD} datasets. In the medical domain, we re-masked the brain medical detection datasets BrainMRI\cite{Ali2025GenerativeAS}, HeadCT\cite{Acosta2024HeadCTONEEG}, and Br35H\cite{tbkk-q937-25}, making them qualified anomaly detection datasets. The experiment follows the Zero-shot setting, our ADSeeker is fine-tuned on MVTec AD to evaluate the ZSAD performance on other datasets. When evaluating the model on MVTec AD, the corresponding train-set is replaced by VisA. To ensure the experiment fairness, we stopped utilizing Q2K RAG to ensure test-set is not leaked during test. We will demonstrate our complete experimental setup and training parameters.

The results are presented in \cref{Tab1}, our ADSeeker have achieved significant enhancement on most of the AD benchmarks in both medical and industrial domains. The enhancement of our model mainly originates from the Hierarchical Sparse Prompt mechanism, which enables the model to extract the defect-region features and learn more fine-grained semantics for anomaly detection in interaction with learnable textual prompts. ADSeeker has emerged as the top performer, boasting the highest average ranking. This achievement highlights its impressive ability to generalize effectively across various domains, as the visualization results shown in \cref{fig6}.

\subsection{Anomaly Reasoning \& ADSeeker Framework}

\begin{table}[t!] 
\centering 
\small
\renewcommand{\arraystretch}{1}
\setlength{\tabcolsep}{4pt} 
\begin{tabular}{lcccc}
\toprule  
\textbf{Model} & \textbf{DS-MVTec} & \textbf{LOCO} & \textbf{VisA} & \textbf{Average} \\
\midrule  
VLM R1\cite{Shen2025VLMR1AS} & 72.01 & 58.47 & 63.61 & 64.69 \\
Anomaly R1\cite{Chao2025AnomalyR1AG} & \underline{81.99} & \underline{70.55} & 63.27 & 71.93 \\
\rowcolor{gray!25} VLM R1* & 77.65 & \textbf{71.03} & \underline{68.11} & \underline{72.26} \\
\midrule  
LLaVA-OV\cite{Li2024LLaVAOneVisionEV} & 70.13 & 63.21 & 59.05 & 64.13 \\
Anomaly OV\cite{Xu2025TowardsZA} & 73.99 & 65.39 & \textbf{70.02} & 69.80 \\
\rowcolor{gray!25} LLaVA-OV* & \textbf{82.77} & 70.31 & 67.55 & \textbf{73.54} \\
\bottomrule 
\end{tabular}
\caption{\textbf{Comparison with training-methods.} Anomaly reasoning performance comparison with and the best-performing training methods. * represents base model under the Seek-Setting.}
\label{tab2}
\end{table}

With the powerful capability of retrieving relevant expertise and the anomaly prior feature from the AD Expert module, ADSeeker has achieved great improvement in performing anomaly reasoning tasks. As is shown in \cref{fig2}, ADSeeker could accurately locate defects and point out the defect type. Its anomaly description could achieve the accuracy of human anomaly inspectors.

We evaluated anomaly reasoning capability via accuracy on multiple-choice subtasks from the MMAD benchmark, comprising four anomaly-related and two object-related subtasks. Under a default 1-shot setting, compared models received a randomly selected normal sample as template alongside the query image. For ADSeeker and Seek-Setting models, the retrieved knowledge document replaced this normal sample as the template. As shown in \cref{tab4}, ADSeeker (i.e., Qwen2.5-VL under Seek-Setting) was compared against open-source MLLMs. And each model is also tested under Seek-Setting equipped with our Q2K RAG and AD Expert module to validate the efficacy of the plug-and-play framework. ADSeeker delivers a remarkable accuracy of 69.90\%, excelling across various subtasks. And our framework shows enhance mainly on defect classification, localization and Anomaly Discrimination. The injected anomaly prior feature enhances defect localization through visual information fusion, while the retrieved knowledge document—containing specific defect types—facilitates defect classification and anomaly discrimination.

Pretraining remains the most common solution for anomaly reasoning. Recent best-performing training-based approaches (e.g., Anomaly-OV and Anomaly-R1) have achieved remarkable results in IAD tasks. To evaluate the efficacy of our framework, we compare their anomaly reasoning performance under both original training strategies and the Seek-Setting. Due to the limitation of SEEK-M\&V content, we utilize part of the MMAD dataset to conduct evaluation. As illustrated in \cref{tab2}, extensive experiments demonstrate that ADSeeker could outperform existing SOTA IAD approaches without additional training. Notably, Anomaly-R1 achieves remarkable performance by adapting the GRPO strategy to fine-tuning with a limited training set, the utilization of that strategy is suggested for future works.


\subsection{Efficiency and Efficacy Analysis}
\begin{table}[]
    \centering
    \small 
    \setlength{\tabcolsep}{4pt} 
    
    \begin{tabular}{ccccc} 
        \toprule
        \multicolumn{3}{c}{\textbf{ADSeeker}} & \multirow{2}{*}{LoRA\cite{Hu2021LoRALA}} & \multirow{2}{*}{\textbf{Accuracy}} \\ 
        \cmidrule(r){1-3} 
        Baseline\cite{Bai2025Qwen25VLTR} & Q2K RAG & Expert & & \\ 
        \midrule
        \ding{51} & & & & (76.6, 67.1) \\
        \ding{51} & \ding{51} & & & (78.6, 68.2) \\
        \ding{51} & & \ding{51} & & (77.9, 67.7) \\
        \rowcolor{gray!25}
        \ding{51} & \ding{51} & \ding{51} & & \textbf{(82.8, 71.4)} \\
        \ding{51} & & & 5 & (80.0, 71.0) \\
        \ding{51} & & & 10 & (72.8, 59.7) \\
        \ding{51} & & & 20 & (40.9, 33.2) \\ 
        \bottomrule
    \end{tabular}
    \caption{\textbf{Ablation Results of proposed modules and training settings.} The \ding{51} in each columns signifies module integration. The number in \enquote{LoRA} column denotes we utilize LoRA to fine-tune MLLM for that epochs.}
    \label{tab3}
\end{table}

\begin{table}[htbp]
\centering
\small 
\setlength{\tabcolsep}{3pt} 
\begin{tabular}{c|ccccc}
\toprule
\textbf{Efficiency} & \textbf{Q2K RAG} & \textbf{Expert} & \textbf{Base Model} & \textbf{ADSeeker} \\ 
\midrule
Mem (GiB)  & 4.25±0.53 & 2.44±0.09 & 22.63±1.35 & 28.36±1.68 \\
Time (s)  & 1.13±0.16 & 1.52±0.02 & 4.47±2.16 & 6.14±2.20 \\
\bottomrule
\end{tabular}
\caption{\textbf{Quantification of reasoning efficiency.} Memory usage and inference time calculation of each module (mean±SD).}
\label{tab5}
\end{table}

To demonstrate the efficacy of each proposed module in ADSeeker, we conduct ablation experiments by comparing the accuracy of ADSeeker in 2 subtasks among MMAD. We primarily focus on 3 aspects: the Q2K RAG, the AD Expert module and the impact of LoRA fine-tuning on ADSeeker. Our team constructed a 43K small-scale instruction tuning dataset, we collected domain knowledge from AD datasets and our SEEK-M\&V content. The dataset content contains most of the knowledge base content and is constructed using template examples. As shown in \cref{tab3}, the full-configuration ADSeeker achieves optimal performance across tasks. Each module contributes uniquely to anomaly reasoning: Q2K RAG retrieves domain knowledge, while the AD Expert module provides anomaly prior features for localization and discrimination. Note that model capability peaks then declines with extended training epochs. And small-scale training sets exacerbate overfitting and catastrophic forgetting of pretrained knowledge. Moreover, our Q2K RAG framework surpasses LoRA fine-tuning in both generalizability and precision.  

As is shown in \cref{tab5}, we calculate the Memory Usage and Inference Time of each module to demonstrate the efficiency of our proposed framework. When deployed on LLMs of comparable scale, ADSeeker incurs modest overhead: $\leq27\%$ additional memory consumption and $\leq2s$  average inference latency. This demonstrates strong potential in real-time industrial applications. Furthermore, ADSeeker is a training-efficient framework, which shows no need for large-scale pre-training and effectively reduces the consumption of computational resources. In general, proposed as a plug-and-play framework, ADSeeker could achieve the trade-offs between performance and computational resources.

\begin{figure}[t!]
    \includegraphics[width=1.02\linewidth]{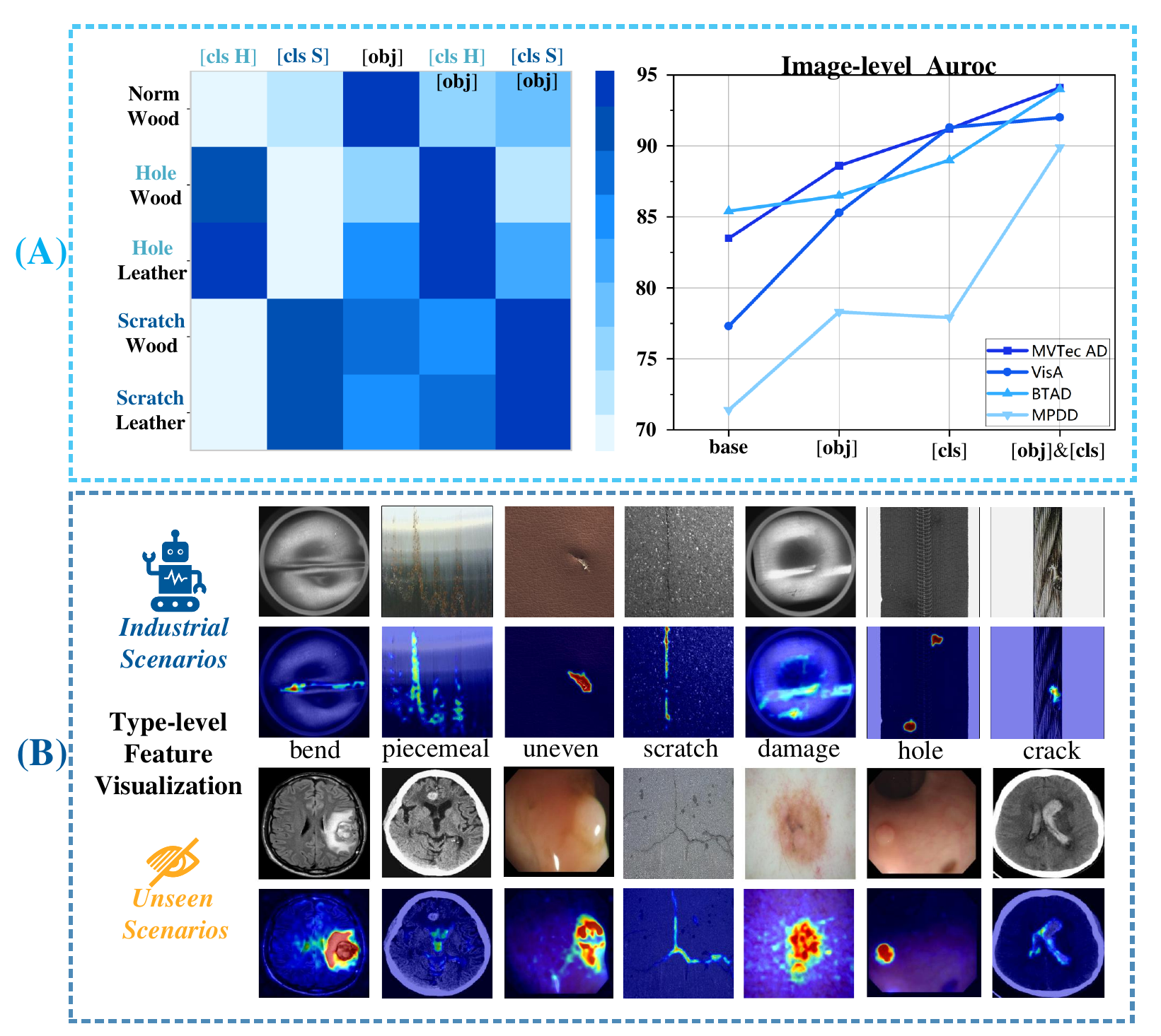}
\caption{\textbf{Validation of type-level feature representation.}}
\label{fig6}
\end{figure}

\subsection{Influence of Type-level Feature}
On the left of the \cref{fig6}(A), we calculate the similarity of the patch tokens and different kinds of textual prompts. Experiments reveal stronger alignment between normal samples and \textbf{[obj]}-only prompts, while anomalous samples exhibit peak similarity to prompts containing both \textbf{[cls]\&[obj]}. Furthermore, for defect samples, [cls] prompts demonstrate significantly higher feature value than [obj] prompts. These findings empirically validate the efficacy of type-level information representation. Our cross-dataset experiments of different-level prompts efficacy reveals that type-level [cls] features significantly outperform [obj] features in image-level anomaly detection tasks. However, [obj] features maintain a noteworthy complementary role, as evidenced by the ablation study in \cref{fig6}.

 As shown in \cref{fig6}(B), we utilize the type-level feature to detect anomaly industry images and provide visualization results. Combining the Image-Level AUROC in \cref{Tab1} and visualization results, we concluded that the ability to extract general features is the key to achieving zero-shot rendering.

\section{Conclusion}
In this paper, we propose an anomaly task assistant ADSeeker, leveraging the visual document Q2K RAG to seek for domain knowledge. To equip ADSeeker with expertise, we establish the first visual document knowledge base \textit{SEEK-M\&V} for MLLMs in IAD tasks. Our work delves into the potential application of external knowledge in anomaly reasoning and detection, introducing a knowledge-grounded reasoning framework without large-scale training.

\section{Acknowledgements}
This work was supported by the National Natural Science Foundation of China (Grant No. U24A20221), the Key R\&D Program of Shandong Province, China (Grant No. 2023CXGC010112), the Distinguished Young Scholar of Shandong Province (Grant No. ZR2023JQ025), the Taishan Scholars Program (Grant No. tstp20250708), and the Major Basic Research Projects of Shandong Province (Grant No. ZR2022ZD32).

{
    \small
    \bibliographystyle{ieeenat_fullname}
    \bibliography{main}
}

\end{document}